\documentstyle[12pt,a41]{article}
\newcommand\XINT{\int_{-\infty}^{+\infty}}
\newcommand\YINT{\int_{-1}^{+1}}
\newcommand{\bq}{\begin{equation}}
\newcommand{\eq}{\end{equation}}

\newcommand\ka{\kappa_1}
\newcommand\kb{\kappa_2}
\newcommand\kap{\kappa_1'}
\newcommand\kbp{\kappa_2'}
\newcommand\xx{\tilde{x}}
\newcommand\AAA{\alpha_1}
\newcommand\AB{\alpha_2}
\newcommand\Kvec{\mbox{\boldmath $K$}}

\newcommand\bx{\overline{x}}
\newcommand\by{\overline{y}}
\newcommand\FFA{\mbox{$\widetilde{F}^a$}}
\newcommand\FA{\mbox{${F}^a$}}

\begin{document}
\sloppy
\thispagestyle{empty}
\begin{flushleft}
DESY 97--209 \hfill {\tt hep-ph/9711405}\\
November 1997
\end{flushleft}
\mbox{}
\vspace*{\fill}
\begin{center}
{\LARGE\bf  Twist-2 Light-Ray Operators: }

\vspace{2mm}
{\LARGE\bf Anomalous Dimensions and Evolution}

\vspace{2mm}
{\LARGE\bf Equations}\footnote{ Contribution to the Proceedings
of the Int. Workshop {\sf Deep Inelastic Scattering off Polarized
Targets : Theory Meets Experiment}, September 1--5, 1997,
DESY--Zeuthen, Germany}
\\

\vspace{2em}
\large
Johannes Bl\"umlein$^a$, Bodo Geyer$^b$, and Dieter
Robaschik$^{a,c}$
\\
\vspace{2em}
{\it $^a$~DESY--Zeuthen, Platanenallee 6,
D--15735 Zeuthen, Germany}\\
\vspace{2em}
{\it $^b$~Naturwissenschaftlich--Theoretisches Zentrum,}\\
{\it
Universit\"at Leipzig,
 Augustusplatz 10, D--04109 Leipzig, Germany}
\\
\vspace{2em}
{\it $^c$~Institut f\"ur
Theoretische Physik der Karl-Franzens-Universit\"at  Graz,}\\
{\it  Universit\"atsplatz 5, A-8010 Graz, Austria} \\
\end{center}
\vspace*{\fill}
%
\begin{abstract}
\noindent
The non--singlet and singlet anomalous dimensions of the twist--2
light--ray operators for unpolarized and polarized deep inelastic
scattering are calculated in $O(\alpha_s)$. We apply these
results for the derivation of evolution equations for partition
functions, structure functions, and wave functions which are defined
as Fourier transforms of the matrix elements of the light-ray operators.
Special cases are the Altarelli--Parisi and Brodsky--Lepage kernels.
Finally we extend Radyushkin's solution from the non--singlet to the
singlet case.
\end{abstract}
\vspace*{\fill}
\newpage
\noindent
\section{Introduction}
The study of the Compton amplitude for scattering a virtual
photon
off a hadron is one of the basic tools in QCD to understand the
short--distance behavior of the theory.
In the general
case of  {\em non--forward} scattering the Compton amplitude is
given by
\begin{equation}
\label{COMP}
T_{\mu\nu}(p_+,p_-,Q) = i \int d^4x e^{iqx}
\langle p_2|T (J_{\mu}(x/2) J_{\nu}(-x/2))|p_1\rangle,
\end{equation}
where $p_\pm = p_2 \pm p_1, Q = (q_1 + q_2)/2,$ and
$p_1 + q_1 = p_2 + q_2$.

The time--ordered product  in eq.~(\ref{COMP}) can be
represented in terms of the {\em non-local} operator product
expansion \cite{AS}~:
\begin{eqnarray}
\label{tpro}
T (J_{\mu}(x/2) J_{\nu}(-x/2)) &\approx&
\int_{-\infty}^{+\infty} d \kappa_-
\int_{-\infty}^{+\infty} d \kappa_+
\left [ C_a(x^2, \kappa_-, \kappa_ +, \mu^2) {S_{\mu\nu}}^{\rho\sigma}
\xx_{\rho} O_{\sigma}^a (\kappa_+ \xx, \kappa_- \xx, \mu^2) \right.
\nonumber\\
& & \left.
~~~~~~~~~~~~~~~~~~~~+
C_{a,5}(x^2, \kappa_-, \kappa_ +, \mu^2)
{\varepsilon_{\mu\nu}}^{\rho\sigma}
\xx_{\rho} O_{5, \sigma}^a (\kappa_+ \xx, \kappa_- \xx, \mu^2) \right ],
\end{eqnarray}
with $
    S_{\mu \nu \lambda \tau} =
    g_{\mu\nu} g_{\lambda \sigma}
    - g_{\mu\lambda} g_{\nu \sigma}
    - g_{\mu\sigma} g_{\lambda \nu} $ and
$\varepsilon_{\mu\nu\rho\sigma}$ denoting the Levi--Civita symbol.
The light--like vector
\begin{equation}
\xx = x - \eta (x.\eta/\eta.\eta)
+ \eta \frac{x.\eta}{\eta.\eta}\sqrt{1 - \frac
{x.x \eta.\eta}{(x.\eta)^2}}
\end{equation}
is related to $x$ and a constant subsidiary four--vector
$\eta$ which drops out in leading order.
Unlike the local operator product expansion~\cite{OPEL}, which
is usually  applied in the case of forward scattering,
eq.~(\ref{tpro})  straightforwardly leads to  compact expressions for the
coefficient functions $C_a(x^2, \kappa_\pm, \mu^2)$ and light--ray
operators $O^{a\sigma}_{(5)}(\kappa_\pm{\tilde x}, \mu^2).$
Indeed this  expansion is a summed--up local
light--cone expansion \cite{AS,SLAC}. In lowest order it
contains only quark operators with two
external legs. These operators are to be decomposed into
operators of different twist. Here, we consider the contribution of
twist--2  operators only\footnote{The notion of twist is not quite
unambiguous in this context. It is rather used to label the type
of operators being concerned, for which in the limit
$p_2 \rightarrow p_1$ the notion applies~\cite{GROSS}.}
for which  we calculate the anomalous dimensions
entering the
corresponding renormalization group equation \cite{BGR1}.

Different processes, such as deep inelastic (forward)
scattering~\cite{AP},
deeply virtual Compton scattering~[5,7--12],
or more generally non--forward
scattering processes, cf. e. g.~\cite{LEIP},
and photoproduction of mesons~[13--16]
contain as an essential input the same light--ray operators of twist--2:
All interesting non--perturbative partition functions are Fourier
transforms of corresponding matrix elements of these operators.
Therefore, all evolution equations result from the renormalization group
equation of the  light--ray operators involved.
The evolution kernels can
be determined from the anomalous dimensions of these operators. This
procedure will be carried out here. The Altarelli--Parisi kernels
as well as the Brodsky-Lepage/Efremov-Radyushkin kernels appear as
special cases.

A second interesting feature of this approach is that non--forward
processes contain more variables than forward processes. In the
generalized Bjorken region this implies that instead of only
one distribution parameter  in the forward case, the Bjorken
variable $x$, two distribution parameters appear in the present case.
One of them
is related to the sum and the other to the difference of the
external momenta. The explicit solution of the evolution equation
in the non--singlet case was firstly found by Radyushkin~\cite{RADL}.
Here, we extend his method to the singlet case. On the other hand,
by imposing specific conditions on the momenta also one-parameter
equations may be obtained  which are equivalent to the general case
treating a parameter which is fixed for special cases as a free second
variable.

In this paper we present the results of a calculation of the
non--singlet
and singlet twist-2
anomalous dimensions for the  general case both for
unpolarized and polarized scattering, cf.~\cite{BGR1}.
We derive general evolution equations both for the operators and
matrix elements in the non--forward case.
Furthermore we derive
a series of special cases previously discussed in the literature.
\section{Twist-2 Light-Ray Operators and Different Choices of
Partition Functions }
We consider the following twist-2 flavor
singlet operators\footnote{In leading order the non--singlet operator
is related to eq.~(\ref{q}). The corresponding results which are derived
below for this operator apply literally to
$O_{\rm NS}(\kappa_1,\kappa_2)$ in this order.}
which are known to mix under renormalization
\begin{eqnarray}
\label{q}
O^q_{\rm     }(\ka,\kb) &=&
\frac{i}{2} \left [
\overline{\psi_r}(\ka\xx)
\gamma_{\mu} \xx^{\mu} \psi_r(\kb \xx) -
\overline{\psi_r}(\kb\xx)
\gamma_{\mu} \xx^{\mu} \psi_r(\ka \xx) \right ],\\
\label{G}
O^G_{\rm     }(\ka,\kb) &=&
\xx^{\mu} \xx^{\mu'}
{F^a}_{\mu}^{\nu}(\ka\xx)  {F^a}_{\mu'\nu}(\kb\xx),\\
\label{q5}
O^q_{\rm 5   }(\ka,\kb) &=&
\frac{i}{2}
\left [
\overline{\psi_r}(\ka\xx) \gamma_5
\gamma_{\mu} \xx^{\mu} \psi_r(\kb \xx) +
\overline{\psi_r}(\kb\xx) \gamma_5
\gamma_{\mu} \xx^{\mu} \psi_r(\ka \xx) \right ],\\
\label{G5}
O^G_{\rm 5   }(\ka,\kb) &=& \frac{1}{2} \xx^{\mu}\xx^{\mu'}
\left [
 {F^a}_{\mu}^{\nu}(\ka\xx) \FFA_{\mu'\nu}(\kb\xx)
- {\FA}_{\mu\nu}(\kb\xx)
     {\FFA}_{\mu'}{\vspace*{-1.2mm}^{\nu}}(\ka\xx)
\right ]~.
\end{eqnarray}
Here $\psi_r$ denotes the quark field,
$F_{\mu\nu}^a$ and $\widetilde{F}_{\mu\nu}^a $ the gluon field
strength and its dual, respectively, and furthermore we use
\begin{equation}
\kappa_1 = \kappa_+ - \kappa_-,~~~~~~~~~\kappa_2 = \kappa_+ +
\kappa_-~.
\end{equation}
To simplify the considerations we apply
the axial gauge, $\tilde{x}_{\mu} A^{\mu} = 0$, by which the
phase factor connecting gauge dependent fields equals to unity.

The operators
(\ref{G}) and (\ref{G5}) appear only in the next order in $\alpha_s$
in the Compton scattering amplitude. However, due to the mixing of
operators  of the
same dimension and twist they emerge in
the renormalization group equations.
Whereas for the determination of the anomalous
dimensions it is sufficient to investigate the operators
(\ref{q}--\ref{G5}), by considering the
 evolution equations for observables
it is necessary to introduce generalized partition functions
related to  the matrix elements of the
operators $O^i, i = q,G$; (the index 5 will be
suppressed in the following).
The expectation values may be represented by either
\begin{eqnarray}
\label{partap}
\frac{\langle p_1|O^{i}|p_2\rangle}{(i\xx p_+)^{h_i}} &=&
e^{-i\kappa_+ \xx p_-}
\int_{-1}^{+1} d z_+
 d z_-
e^{-i\kappa_-( \xx p_+ z_+ + \xx p_- z_-)} F_{i}(z_+,z_-)
\end{eqnarray}
or
\begin{eqnarray}
\label{partrad}
\langle p_1|O^{i}|p_2\rangle \cdot (\kappa_-)^{h_i} &=&
e^{-i\kappa_+ \xx p_-}
\int_{-1}^{+1} d z_+
 d z_-
e^{-i\kappa_-( \xx p_+ z_+ + \xx p_- z_-)} G_{i}(z_+,z_-)~.
\end{eqnarray}
Here $h_i$ denotes
the degree of
homogeneity, $h_i$, with respect to a rescaling of the $\kappa$-variables
of the quarkonic and gluonic operators  (cf. \cite{BGR}) with
\begin{equation}
\label{degree}
h_q = 1,~~~~~~~~~~~~~~~~~~h_G = 2~.
\end{equation}
One way to express the partition function, eq.~(\ref{partap}),
is connected with the {\em states} and
has the property, that in the forward
limit, $|p_1\rangle \rightarrow |p_2\rangle$,
these amplitudes lead to the well known structure functions
appearing in the Altarelli--Parisi-equations.
The second expression, eq.~(\ref{partrad}),  which is related
to the light--like spread of the {\em operators} leads to
simpler two--variable evolution equations  the solution of which is
more straightforwardly obtained, cf.~section~7.
\section{Anomalous Dimensions of Twist 2 Operators  }
The renormalization group equations for the singlet operators read:
\begin{eqnarray}
\label{evo2}
\mu^2 \frac{d}{d \mu^2}
O^i(\ka,\kb)
&= &
\frac{\alpha_s(\mu^2)}{2 \pi}
\int D \alpha
K^{ij}(\AAA,\AB,\kappa_-)
O_j(\kap,\kbp)
\end{eqnarray}
with
\begin{eqnarray}
 D\alpha  =    d \AAA
d \AB~\theta(1 - \AAA - \AB)~\theta(\AAA)~\theta(\AB),
\end{eqnarray}
$\alpha_s = g_s^2/(4\pi)$ the strong coupling constant,
$\mu$ the renormalization scale, and
\begin{eqnarray}
\kap  =  (1-\AAA)\ka + \AAA \kb, \; \; \;
\kbp  =  \AB \ka + (1-\AB) \kb.
\nonumber
\end{eqnarray}
By $\Kvec = (K^{ij})$
we denote the
matrix of singlet
anomalous dimensions in  the unpolarized case. The singlet
evolution
equations for the polarized case are obtained by replacing the operators
$O^{q,G}$ by
$O^{q,G}_{5}$ and the matrix $\Kvec$ by $\Delta \Kvec$.
As far as relations are
concerned which are valid both for the unpolarized and the polarized
case
under this replacement, we will, for brevity, only give that for
the
unpolarized case  in the following. The matrices of the
singlet anomalous dimensions  are
\begin{equation}
\Kvec = \left ( \begin{array}{ll} K^{qq} & K^{qG} \\
                                  K^{Gq} & K^{GG} \end{array}
\right )
~~~~{\rm and}~~~~\Delta \Kvec = \left ( \begin{array}{ll} \Delta
K^{qq} &
\Delta K^{qG} \\ \Delta K^{Gq} & \Delta K^{GG} \end{array}
\right )~,
\end{equation}
respectively.  In leading order, the
non--singlet anomalous dimensions obey
$K^{\rm NS} = \Delta K^{\rm NS} = K^{qq} = \Delta K^{qq}$.

For the unpolarized case the anomalous dimensions are given by
\begin{eqnarray}
\label{eqK1}
K^{qq}(\AAA,\AB) &=&  C_F \left \{ 1 - \delta(\AAA) -
\delta(\AB)
+ \delta(\AAA) \left [ \frac{1}{\AB}\right]_+
+ \delta(\AB) \left [ \frac{1}{\AAA}\right]_+
+ \frac{3}{2} \delta(\AAA)\delta(\AB) \right \},\\
K^{qG}(\AAA,\AB) &=&  - 2 N_f T_R \kappa_-
\left \{ 1 - \AAA - \AB + 4 \AAA\AB
\right \}, \\
K^{Gq}(\AAA,\AB) &=&  - C_F \frac{1}{\kappa_- }
\left \{  \delta(\AAA) \delta(\AB) + 2 \right \}, \\
\label{eqK4}
K^{GG}(\AAA,\AB) &=&  C_A \left \{ 4 ( 1 - \AAA -\AB) + 12 \AAA
\AB
\right.
\\  & &~~+ \left.
\delta(\AAA) \left (
 \left [\frac{1}{\AB} \right ]_+         - 2
+ \AB \right )
+ \delta(\AB) \left ( \left [\frac{1}{\AAA} \right ]_ +
- 2
+ \AAA \right )  \right \}
+ \frac{\beta_0}{2} \delta(\AAA)\delta(\AB),   \nonumber
\end{eqnarray}
in ${O}(\alpha_s)$
where $C_F = (N_c^2-1)/2N_c \equiv 4/3, T_R = 1/2, C_A = N_c
\equiv 3$,
$\beta_0 = (11 C_A - 4 T_R N_f)/3$.
The $[~]_+$--prescription is defined by
\begin{equation}
\label{+pres}
\int_0^1 dx
 \left [f(x,y)\right]_+ \varphi(x) = \int_0^1 dx
f(x,y)
\left [\varphi(x) - \varphi(y) \right]~,
\end{equation}
if  the singularity of $f$ is of the type $\sim 1/(x - y)$.

Correspondingly
the  anomalous dimensions  for the polarized case are
\begin{eqnarray}
\label{eqDK1}
\Delta K^{qq}(\AAA,\AB) &=& K^{qq}(\AAA,\AB), \\
\Delta K^{qG}(\AAA,\AB) &=&  - 2 N_f T_R \kappa_-
\left \{ 1 - \AAA - \AB \right \}, \\
\Delta K^{Gq}(\AAA,\AB) &=&  -  C_F \frac{1}{\kappa_- }
\left \{\delta(\AAA) \delta(\AB) - 2 \right \}, \\
\label{eqDK4}
\Delta K^{GG}(\AAA,\AB) &=&  K^{GG}(\AAA,\AB) - 12 C_A   \AAA
\AB~.
\end{eqnarray}
Whereas the anomalous dimensions for
 the polarized case are derived for the first time in
\cite{BGR1}\footnote{A different, but related, quantity was studied in
\cite{BRAUN}.}, those for the unpolarized case were found several years
ago \cite{BGR,BB}. Recent calculations have  also been performed
in \cite{RAD,RADB}. The matrices
$K^{ij}$ and $\Delta K^{ij}$ determine the
evolution  of the operators 
$O^{\rm NS}_{(5)}, O^{q}_{(5)}$, and $O^{G}_{(5)}$, respectively,
in $O(\alpha_s)$.

The determination of these anomalous dimensions is
straightforward.
In lowest order one has to calculate the one-particle
irreducible
one-loop Feynman diagrams containing the operators considered as
first vertex. The corresponding Feynman rules and the calculation
in the
covariant gauge for one example
are presented in ref.~\cite{BGR}.
Here we have performed the calculation in the
axial gauge which
leads to essential simplifications.  We used the
Leibbrandt-Mandelstam prescription for the axial poles~\cite{LEIB}.
\section{Evolution Equations of Partition Functions}
The  non--perturbative partition functions defined in section~2
have to be determined experimentally at an input scale
$Q_0^2$.
Their values at higher scales $Q^2$ can be calculated perturbatively
solving the corresponding evolution equations.
Here we derive only the equations which emerge from our
two  choices eqs.~(\ref{partap},\ref{partrad}).
These equations follow immediately from the renormalization group
equations of the operators, eq.~(\ref{evo2}).

For simplicity we introduce different
variables which are most convenient, cf.~\cite{BGR}, for the subsequent
discussion
\begin{eqnarray}
\label{w}
w_1 &=& \AAA -\AB = \frac{{\kappa_+}' - \kappa_+}{\kappa_-},
\quad
w_2 = 1-\AAA -\AB = \frac{{\kappa_-}'}{\kappa_-}, \\
\label{dw}
D\alpha &=&
\hbox{$\frac{1}{2}$} Dw = \hbox{$\frac{1}{2}$} dw_1 dw_2
\Theta(1-w_1-w_2) \Theta(1+w_1-w_2)\Theta(w_2).
\nonumber
\end{eqnarray}
In these variables the RG-equation (\ref{evo2})  reads
\begin{eqnarray}
\label{evo2e}
\mu^2 \frac{d}{d \mu^2}
O^i(\kappa_+,\kappa_-)=
\frac{\alpha_s(\mu^2)}{2 \pi}
\int Dw
K^{ij}(w_1,w_2,\kappa_-)
O^j(\kappa_+ + w_1 \kappa_-,w_2 \kappa_-).
\end{eqnarray}
By Fourier transform we get an
evolution equation for the distribution functions
$G^i$, inserting  the definition
(\ref{partrad}) also into the right hand side
\begin{eqnarray}
\mu^2 \frac{d}{d \mu^2}G^i(z_+,z_-) =
 \XINT
 \frac{d\kappa_- \tilde x p_+}{2\pi}
 \XINT \frac{d\kappa_- \tilde x p_-}{2\pi}
e^{i(\kappa_+ - {\kappa_+}')\tilde x p_-}
e^{i\kappa_-(\tilde x p_-z_- +\tilde x p_+ z_+)}
(\kappa_-)^{h_i}
\nonumber \\
~~~~~~~~~~~\frac{\alpha_s(\mu^2)}{4 \pi}
\int Dw K^{ij}(w_1,w_2,\kappa_-)\YINT \YINT
 d{z'_-}d{z'_+}
(\kappa'_-)^{-h_j}
e^{-i\kappa_-'(\tilde x p_+ z'_+ +\tilde x p_- z'_-)}
 G^j({z'_+}{z'_-}).\nonumber
\end{eqnarray}
The $\kappa_-$--dependence of the
evolution kernel is $K^{ij}(w_1,w_2,\kappa_-) =
\hat K^{ij}(w_1,w_2) (\kappa_-)^{h_j - h_i}$. Carrying out the
integrations with respect to
$d\kappa_- \tilde x p_{\pm}$ we get
\begin{eqnarray}
\label{1}
\mu^2 \frac{d}{d \mu^2}G^i(z_+,z_-)&=&
\frac{\alpha_s(\mu^2)}{4 \pi}
\int Dw \hat K^{ij}(w_1,w_2)
w_2^{-h_j}
\YINT d{z'_-} \YINT
d{z'_+}\nonumber \\
& &  \delta(z_- - w_2 {z'_-}-w_1)
\delta(z_+ - w_2 {z'_+})
 G^j({z'_+}{z'_-})
\end{eqnarray}
leading to
\begin{eqnarray}
\label{evo22}
\mu^2 \frac{d}{d \mu^2}G^i(z_+,z_-)=
 \frac{\alpha_s(\mu^2)}{4 \pi}
\YINT \YINT
\frac{dz'_- dz'_+}{|z'_+|}
\hat K^{ij}
     (z_- -\frac{z_+}{z'_+} z'_-, \frac{z_+}{z'_+})
     \left(\frac{z'_+}{z_+}\right)^{h_j} G^j({z'_+}{z'_-})~.
\end{eqnarray}
The derivation of the evolution equations for the partition
functions
(\ref{partap}) follows the same line.
The difference is that the
integration with respect to $d\kappa_- \tilde x p_+ $ leads {\it not}
directly
to a $\delta$-function so that one obtains
\begin{eqnarray}
\mu^2 \frac{d}{d \mu^2}F^i(z_+,z_-)=
 \frac{\alpha_s(\mu^2)}{2 \pi}
\YINT \YINT
 dz'_- dz'_+  \Gamma^{ij}(z_+, z_-;z'_+,z'_-)
 F^j({z'_+},{z'_-}),
\end{eqnarray}
where
\begin{eqnarray}
 \Gamma^{ij}(z_+,z_-,z'_+,z'_-) = \int
 \frac{dw_2}{2}\XINT \frac{d\kappa_-\tilde x p_+}
{4\pi}\frac{e^{i\kappa_- \tilde x p_+ (z_+ - w_2 z'_+)}}
{(i\kappa_-\tilde x p_+)^{h_{ij}}}
\hat K^{ij}(z_- -w_2 z'_-,w_2)~,
\end{eqnarray}
with $h_{ij} = h_i - h_j$.
Wether the $d\kappa_- \tilde x p_+ $--integration leads to a
$\delta$--function, its derivative, or an integration depends on
the values
of $i$ and $j$. In the case of an integration a
$\Theta$--function appears, for which the integration constant has
to be determined separately.
\section{One-Variable Evolution Equations}
%
The equations given above cover the most general case depending on
two partition variables $z_+$ and $z_-$. To reveal the relation of the
general case to more special cases we introduce a kinematic condition
by $\tilde x p_- = \tau \tilde x p_+ $. This condition appears natural
having in mind scale invariance. For special cases being discussed
below the parameter $\tau$ will be
fixed. On the other hand, one may consider
$\tau$ as a second {\it general}
 parameter aside $\tilde x p_+$ resembling the
general two--variable case.
Instead of the variables $z_-$ and $z_+$ we use  $t$ and
$\tau$ as new variables defining
\begin{eqnarray}
  \Phi_i(t,\tau)& = &\XINT dz_- F_i(t - \tau z_-, z_-), \\
  F_i(z_+,z_-) &=&\frac{1}{(2\pi)^2}
 \XINT dt \XINT
d\tau \frac{|{z}_-|}{\tau^2}
                e^{i{{z}_-}(z_+ - t)/\tau
}\Phi_i(t,\tau)~.
\end{eqnarray}
These partition functions are related to the expectation values
of the operators $Q^q$ and $O^G$ by
\begin{eqnarray}
\left.
\frac{
\langle p_1|O^{q}(-\kappa_- \xx,
\kappa_- \xx)|p_2\rangle}
{(i\xx p_+)}
\right|_{\xx p_- = \tau \xx p_+}
&=&
\int_{-\infty}^{+\infty} dt
e^{-i\kappa_- \xx p_+ t} \Phi_{q}(t,\tau),\\
\left.
\frac{
\langle p_1|O^{G}(-\kappa_- \xx,
\kappa_- \xx)|p_2\rangle}
{(i\xx p_+)^2}
\right|_{\xx p_- = \tau \xx p_+}
&=&
\int_{-\infty}^{+\infty} dt
e^{-i\kappa_- \xx p_+ t}~t \Phi_{G}(t,\tau)~.
\end{eqnarray}
The evolution equations are~:
\begin{eqnarray}
\label{evo3ta}
\mu^2 \frac{d}{d \mu^2}
\Phi^{i}(t,\tau)
 &=&
\frac{\alpha_s(\mu^2)}{2 \pi}
\int_{-\infty}^{+\infty} d t'
V^{ij}_{ext}(t,t',\tau)
\Phi^{j}(t',\tau).
\end{eqnarray}
The corresponding extended kernels
read
\begin{eqnarray}
\label{ker3t}
V_{ext}^{ij}(t,t',\tau) &=&
\int D \alpha
K^{ij}(\AAA,\AB,\kappa_-)
\frac{1}{2 \pi} \int_{- \infty}^{+ \infty} d(p_+\xx \kappa_-)
\left[i p_+\xx 
\right]^{h_{ij}}\\  & \times &
\frac{{t'}^{h_j - 1}}{t^{h_i - 1}}
\exp\left
\{i p_+ \kappa_- \xx\left[t-(1-\AAA -\AB)t' + \tau (\AAA
-\AB)\right]
\right\}  \nonumber
\end{eqnarray}
and obey the scaling relation
\begin{eqnarray}
V^{ij}_{ext}(t,t',\tau) = \frac{1}{\tau}
V_{ext}^{ij}\left(\frac{t}{\tau},
\frac{t'}{\tau},1 \right)~.
\end{eqnarray}
For convenience we write the
general expressions for the evolution kernels in the variables
\begin{eqnarray}
\label{xy}
x= \frac{1}{2}\left ( 1 + \frac{t}{\tau}\right),~
\overline x= \frac{1}{2}\left ( 1 - \frac{t}{\tau}\right),~
y= \frac{1}{2}\left ( 1 + \frac{t'}{\tau}\right),~
\overline y= \frac{1}{2}\left ( 1 - \frac{t'}{\tau}\right)~.
\end{eqnarray}
They are given by~:
\begin{eqnarray}
\label{kerGE1}
V^{qq}_{ext}(t,t',\tau) &=&
\frac{1}{2\tau}
\left \{  V^{qq}(x,y)\rho(x,y) + V^{qq}(\bx,\by) \rho(\bx,\by)
+ \hbox{$\frac{3}{2}$} C_{F} \delta(x-y)
\right \}
\\
V^{qG}_{ext}(t,t',\tau) &=&\frac{1}{2\tau}
\left \{  V^{qG}(x,y)\rho(x,y) - V^{qG}(\bx,\by) \rho(\bx,\by)
\right \}
\left(\frac{2y -1}{2} \right)
\\
V^{Gq}_{ext}(t,t',\tau) &=&\frac{1}{2\tau}
\left \{  V^{Gq}(x,y)\rho(x,y)
- \overline{V}^{Gq}(\bx,\by) \rho(\bx,\by)
\right \}
\left( \frac{2}{2x - 1} \right)
\\
\label{kerGE4}
V^{GG}_{ext}(t,t',\tau) &=& \frac{1}{2\tau}
\left \{
V^{GG}(x,y)\rho(x,y) + V^{GG}(\bx,\by) \rho(\bx,\by) \right \}
\left(\frac{2y - 1}{2x - 1}  \right)
\\
& &+ \frac{1}{4\tau} \beta_0 \delta(x-y)
\nonumber\\
\label{kerDGE1}
\Delta V^{qq}_{ext}(t,t',\tau) &=&  V^{qq}_{ext}(t,t',\tau)\\
\Delta V^{qG}_{ext}(t,t',\tau) &=&\frac{1}{2\tau}
\left \{\Delta  V^{qG}(x,y)\rho(x,y) - \Delta V^{qG}(\bx,\by)
\rho(\bx,\by)
\right \}
\left(\frac{2y -1}{2} \right)
\\
\Delta V^{Gq}_{ext}(t,t',\tau) &=&\frac{1}{2\tau}
\left \{\Delta  V^{Gq}(x,y)\rho(x,y)
- \Delta \overline{V}^{Gq}(\bx,\by)
\rho(\bx,\by)
\right \}
\left (\frac{2}{2x - 1} \right )
\\
\label{kerDGE4}
\Delta
V^{GG}_{ext}(t,t',\tau) &=& \frac{1}{2\tau}
\left \{
\Delta V^{GG}(x,y)\rho(x,y) + \Delta V^{GG}(\bx,\by)
\rho(\bx,\by) \right \}
\left( \frac{2y - 1}{2x - 1}  \right)
 \\
& &+  \frac{1}{4\tau} \beta_0 \delta(x-y),\nonumber
\end{eqnarray}
with
\begin{equation}
\rho(x,y) =
\theta\left(1-\frac{x}{y}\right)
\theta\left(\frac{x}{y}\right)~{\rm sign}(y)~,
\end{equation}
and
\begin{eqnarray}
V^{qq}(x,y) &=& C_F
\left [ \frac{x}{y} - \frac{1}{y} + \frac{1}{(y-x)_+}
 \right ]\\
V^{qG}(x,y) &=& - 2 N_f T_R \frac{x}{y} \left [ 4(1 - x) +
\frac{1 - 2x}{y} \right ] \\
V^{Gq}(x,y) &=&  C_F \left [1-  \frac{x^2}{y} \right ] \\
V^{GG}(x,y) &=&  C_A
  \left[2\frac{x^2}{y}\left(3-2x + \frac{1-x}{y}\right)
+\frac{1}{(y-x)_+}
   - \frac{y+x}{y^2}\right] \\
\Delta V^{qq}(x,y) &=& V^{qq}(x,y)\\
\Delta V^{qG}(x,y) &=& - 2 N_f T_R \frac{x}{y^2} \\
\Delta V^{Gq}(x,y) &=& C_F \left [\frac{x^2}{y} \right ] \\
\label{DGE6}
\Delta V^{GG}(x,y) &=& C_A
 \left[2\frac{x^2}{y^2}  +\frac{1}{(y-x)_+}
   - \frac{y+x}{y^2}\right]~.
\end{eqnarray}
Note that the kernels given in
eqs.~(\ref{kerGE1}--\ref{DGE6}) apply to the {\it full} range
of the
variables, i.e. they represent the kernels completely.
The function $V^{qq}_{ext}(t,t',\tau)$ was already derived in
refs.~\cite{ROBA,LEIP}, the complete treatment of both singlet
cases are given in \cite{BGR1}.
For the calculation one may fix
the parameter
 $\tau$, apply the
scaling properties and take into account the necessary general
structure in the $t,~t'$-plane derived in \cite{LEIP}.
Note, that our kernels are represented in a very compact form.
Equivalent
results obtained by other authors~[9--11]
consist at least of two
separately calculated expressions.
For $\tau = 1$, our expressions correspond to $\zeta = 1$ in \cite{RAD}.
In the notation of ref. \cite{XJ}
our parameter $ \tau$  equals to $- \xi/2$. Note also refs.~\cite{OTH}.
\section{Special Cases}
The evolution kernels given above cover  limiting
cases
which were studied before. These are characterized by
special kinematic conditions for the matrix elements.
This concerns the case of forward scattering
$\langle p_2| \rightarrow \langle p_1| \equiv \langle p|$
and the transition from the vacuum state $\langle 0|$ to a hadron
state $\langle p|$  wave functions.
\subsection{The Brodsky--Lepage Limit}
For $\tau = \pm 1$ the equations~(\ref{kerGE1}--\ref{DGE6})
correspond to the limit $\langle p_2| \rightarrow \langle p|,
\langle p_1| \rightarrow \langle 0|$, which is known as the
Brodsky--Lepage~\cite{BL} and Efremov--Radyushkin~\cite{ER}
case.
This limit  $p_1 \rightarrow 0$ may be performed {\it formally},
i.e. irrespective
of any other quantum numbers,
 leading to correct results, cf.~\cite{ROBA}.
The corresponding evolution equations are
given by eq.~(\ref{evo3ta}) using as
variables $x$ and $y$, given in eq.~(\ref{xy}).
As an example we consider the simplest case in the range
$x,y<1, x \neq y$
\begin{eqnarray}
V^{qq}(x,y) =C_F\{\Theta(y-x)[\frac{x}{y} -\frac{1}{y}
+\frac{1}{(y-x)}]
+\Theta(x-y)[
\frac{1-x}{1-y}
-\frac{1}{1-y}
+\frac{1}{(x-y)}]\}.
\end{eqnarray}
\subsection{The Near Forward Representation}
This representation has been introduced in \cite{XJ} and also  used
in ref.~\cite{RAD}. Essentially it consists of the part for
$ t > \tau $ and $ t' > \tau $  of the general kernel.
It
contains the forward scattering evolution kernels
case as limiting cases.
Especially, for a correct
application of the evolution equation to the near forward
matrix elements
this representation is needed (for $t>\tau$ ) but additionally
another representation for $t<\tau$ has to be taken into
account.
In an example we  show  that this representation
follows from our general kernel, i.e.
that the general structure of $V_{ext}^{qq}$
covers also the
case $t>\tau,t'>\tau$. For this range
${\rm sign}(\overline y) = -{\rm sign}(y)$ and
$\Theta(1-\frac{\overline x}
{\overline y}) = \Theta (y-x)$ holds. Using these changes
 and dropping, for simplicity,  the $+$-prescriptions we obtain
\begin{eqnarray}
V^{qq}(x,y) &=&C_F\Theta(y-x)\{\frac{x}{y}[1+\frac{1}{y-x}]
-\frac{1-x}{1-y}[1+\frac{1}{x-y}]\}\nonumber \\
&= &C_F\Theta(y-x)\frac{1}{y-x}
[1 + \frac{x\overline x}{y \overline y}],
\end{eqnarray}
where again $x$ and $y$ are taken from eq.~(\ref{xy})~.
\subsection{The Altarelli--Parisi Limit}
The Altarelli-Parisi kernels can be directly determined from
the anomalous dimensions and suitably defined quark and gluon
distribution functions
\begin{eqnarray}
f^q(z,\mu) &=& \frac{1}{2 \pi} \int_{- \infty}^{+ \infty} d(2p\xx \kappa_-)
\langle p|O^q|p\rangle(\kappa_-,\mu)
\frac{e^{2i p\xx \kappa_- z}}{2ip\xx},
\\
zf^G(z,\mu) &=& \frac{1}{2 \pi}
\int_{- \infty}^{+ \infty} d(2p\xx \kappa_-)
\langle p|O^G|p\rangle(\kappa_-,\mu)
\frac{e^{2i p\xx \kappa_- z}}{(2ip\xx)^2}~.
\end{eqnarray}
The respective polarized parton densities are obtained by replacing
$f^{q,G}(z)$ by $\Delta f^{q,G}(z)$ and $O^{q,G}$ by $O^{q,G}_5$.
In ref.~\cite{BGR1} this procedure has been described and leads to the
known results, \cite{AP}. On the other hand
we should remark, that our general expressions for the
one--variable evolution equations und the corresponding kernels
contain the Altarelli-Parisi case in the limit $\tau \rightarrow 0$.
The determination of this limit is nontrivial because cancellations
of divergent terms occur. As an example
we formulate the basic steps for the
kernel $V_{ext}^{Gq}$.
\begin{eqnarray}
 (2x-1)V_{ext}^{Gq}(x,y)&=&  C_F\{
 \Theta(\frac{x}{y})\Theta(1-\frac{x}{y}) {\rm sign}(y)(1-\frac{x^2}{y})
 \nonumber \\
& & -\Theta(\frac{\overline x}{\overline y})
\Theta(1-\frac{\overline x}{\overline y}) {\rm sign}(\overline y)
(1-\frac{{\overline x}^2}{\overline y})\}\frac{1}{\tau}.
\end{eqnarray}
We re--express the variables
$x,y$ again by the  $t,t'$-variables. Finally the limit $\tau \rightarrow
0$ is performed yielding
\begin{eqnarray}
 \lim_{\tau \rightarrow 0} V_{ext}^{Gq}(t,t')= \frac{1}{t'}
 C_F(\frac{1+(1-z)^2}{z}),~~~~~~z=\frac{t}{t'}.
\end{eqnarray}
This is the corresponding Altarelli-Parisi splitting function.
Another possibility consists in deriving
first the near-forward representation and performing then
the limit to the forward case, cf.~\cite{BGR1}.
\section{Solutions}
Here we look for solutions of the two variable equations in
the singlet case. We extend the method of ref.~\cite{RADL}
to this case.
The idea follows the original solution of the Brodsky-Lepage
equation given by Efremov and Radyushkin \cite{ER}.
The method consists in forming moments of the partition
functions such that the resulting evolution equations for these
moments can be solved.
We start form
eq.~(\ref{evo22}) and look for equations for the moments of
the partition functions (\ref{partrad}),
\begin{eqnarray}
 G_n^i(z_-) = \YINT dz_+ {z_+}^n G^i(z_+,z_-)~.
\end{eqnarray}
One obtains from eq.~(\ref{evo22})
\begin{eqnarray}
\label{evo22m}
& &\mu^2 \frac{d}{d \mu^2}
\YINT dz_+ {z_+}^n G^i(z_+,z_-)\nonumber \\
&=&\YINT dz_+ {z_+}^n\frac{\alpha_s(\mu^2)}{2 \pi}
\frac{1}{2}\YINT d{z'}_-
\YINT \frac{d z'_+}{|z'_+|}\left
(\frac{z'_+}{z_+}\right)^{h_j}
\hat K^{i,j}\left(z_--\frac{z_+}{z'_+}{z'}_-
,\frac{z_+}{z'_+}\right) G^{ij}(z'_+,z'_-) \\
&=& \frac{1}{2}\frac{\alpha_s(\mu^2)}{2 \pi}
\YINT \frac{d z_+}{|z'_+|}\left(\frac{z_+}{z'_+}\right)^{n-h_j}
 \YINT d z'_-
\YINT d z'_+ {z'_+}^n
\hat K^{i,j}\left(z_--\frac{z_+}{z'_+}{z'}_-,\frac{z_+}{z'_+}\right)
 G^{ij}(z'_+,z'_-)
\end{eqnarray}
yielding
\begin{eqnarray}
\label{gne}
\mu^2 \frac{d}{d \mu^2}G_n^i(z_-)=
 \frac{1}{2}\frac{\alpha_s(\mu^2)}{2 \pi}
\YINT dz'_- {\Gamma}_n^{i,j}(z_-,z'_-)G_n^j(z'_-),
\end{eqnarray}
with
\begin{eqnarray}
\label{gan}
 {\Gamma}_n^{i,j}(z_-,z'_-)= \YINT d\frac{z_+}{z'_+}
\left(\frac{z_+}{z'_+}\right)^{n-h_j}
\hat K^{i,j}(z_--\frac{z_+}{z'_+}z'_-,\frac{z_+}{z'_+})~.
\end{eqnarray}
In this way a first diagonalization is obtained.
The kernels ${\Gamma}_n^{i,j}(z_-,z'_-)$ for the unpolarized case read
\begin{eqnarray}
\label{ganqq}
\Gamma_n^{qq} (z,z')& =& C_F\{\Theta(z-z')[(\frac{1-z}{1-z'})^n
 (\frac{1}{n} + \frac{2}{(z-z')_+})]\nonumber \\
& &+\Theta(z'-z)[(\frac{1+z}{1+z'})^n
 (\frac{1}{n} + \frac{2}{(z'-z)_+})]+ 3 \delta(z-z')\},\\
\label{ganqG}
\Gamma_n^{qG} (z,z')& = &-2N_f T_R \{\Theta(z-z')
[(\frac{1-z}{1-z'})^n \frac{n^2 +2n(z-z') -(2zz'-1)}
{(n+1)n(n-1)} ]\nonumber \\
& & + \Theta(z'-z)[(\frac{1+z}{1+z'})^n
 \frac{n^2 +2n(z'-z) -(2zz'-1)}
{(n+1)n(n-1)} ]\},\\
\label{ganGq}
\Gamma_n^{Gq} (z,z')& =&-C_F\{\Theta(z-z')[(\frac{1-z}{1-z'})^n
 \frac{1}{n} ]\nonumber \\
& &+\Theta(z'-z)[(\frac{1+z}{1+z'})^n
 \frac{1}{n} ]+  \delta(z-z')\}, \\
\label{ganGG}
\Gamma_n^{GG} (z,z')& = & C_A \{\Theta(z-z')
[(\frac{1-z}{1-z'})^n
(\frac{3(1+z)(1-z')}{n-1}
+\frac{6zz'}{n}
\frac{3(1+z')(1-z)}{n+1} -\frac{2}{n} +\frac{2}{(z-z')_+}]
\nonumber \\
& & \Theta(z'-z)
[(\frac{1+z}{1+z'})^n
(\frac{3(1-z)(1+z')}{n-1}
+\frac{6zz'}{n}
\frac{3(1-z')(1+z)}{n+1} -\frac{2}{n} +\frac{2}{(z'-z)_+}]
\} \nonumber \\
& & + \beta_0 \delta(z-z').
\end{eqnarray}
Similarly one obtains for the polarized kernels
\begin{eqnarray}
\label{pganqq}
\Delta\Gamma_n^{qq} (z,z')& =& C_F\{\Theta(z-z')
[(\frac{1-z}{1-z'})^n
 (\frac{1}{n} + \frac{2}{(z-z')_+})]\nonumber \\
& &+\Theta(z'-z)[(\frac{1+z}{1+z'})^n
 (\frac{1}{n} + \frac{2}{(z'-z)_+})]+ 3 \delta(z-z')\},\\
\label{pganqG}
\Delta\Gamma_n^{qG} (z,z')& = &-2N_f T_R \{\Theta(z-z')
[\frac{1-z}{1-z'})^n \frac{1}{n} \nonumber \\
& & + \Theta(z'-z)[(\frac{1+z}{1+z'})^n
 \frac{1}{n} ]\},\\
\label{pganGq}
\Delta\Gamma_n^{Gq} (z,z')&
=&+C_F\{\Theta(z-z')[(\frac{1-z}{1-z'})^n
 \frac{1}{n} ]\nonumber \\
& &+\Theta(z'-z)[(\frac{1+z}{1+z'})^n
 \frac{1}{n} ] -  \delta(z-z')\}, \\
\label{pganGG}
\Delta\Gamma_n^{GG} (z,z')& = & C_A \{\Theta(z-z')
[(\frac{1-z}{1-z'})^n
(\frac{4}{n} +\frac{2}{(z-z')_+})]
\nonumber \\
& & \Theta(z'-z)
[(\frac{1+z}{1+z'})^n
(\frac{4}{n} +\frac{2}{(z'-z)_+})]\}
 + \beta_0 \delta(z-z').
\end{eqnarray}
As a
next step we have to find a solution of eq.~(\ref{gan}).
One finds that
the following symmetry--relations hold for kernels
$\Gamma_n^{GG} (z,z')$ and $\Delta\Gamma_n^{GG} (z,z')$~:
\begin{eqnarray}
\label{sym}
   (1-z'^2)^n\Gamma_n^{i,j}(z,z')
   &=&\Gamma_n^{i,j}(z',z)(1-z^2)^n ,\\
   (1-z'^2)^n\Delta\Gamma_n^{i,j}(z,z')
   &=&\Delta\Gamma_n^{i,j}(z',z)(1-z^2)^n~.
\end{eqnarray}
This is expected because of similar problems studied
previously~\cite{RADL,ER}.
The final diagonalization of eq.~(\ref{gne}), referring to the
partition functions eq.~(\ref{partrad})
 for which the quarkonic and gluonic operators are dealt with equally,
can thus be performed using
Gegenbauer polynomials.
\begin{eqnarray}
\label{exp}
  \Gamma_n^{i,j}(z,z')&=& \sum \Gamma^{i,j}_{n,m} (1-z^2)^n
  C^{n+1/2}_m (z) C^{n+1/2}_m (z')N_{n,m} ,\\
\label{pexp}
  \Delta\Gamma_n^{i,j}(z,z')&= &\sum \Delta \Gamma^{i,j}_{n,m}
(1-z^2)^n
   C^{n+1/2}_m (z) C^{n+1/2}_m (z')N_{n,m}~.
\end{eqnarray}
The coefficents $N_{n,m}$  are the normalization factors of the
Gegenbauer polynomials
\begin{equation}
\int_{-1}^{+1} dz (1-z^2)^n C_l^{n+1/2}(z) C_k^{n+1/2}(z) =
\delta_{lk} N_{n,k} = \delta_{lk} \left [
2^{2n+1}
\Gamma^2\left( n + \frac{1}{2}\right) \frac{(k+n+1/2) \cdot
k!}{2 \pi
\Gamma(2n+1+k)}\right]^{-1}~.
\end{equation}
$\Gamma^{i,j}_{n,m}$ and $\Delta \Gamma^{i,j}_{n,m}$
are the respective expansion coefficients which can be easily calculated.
As well-known~\cite{POLY} the various possible representations of
the solution of evolution equations using expansions in orthogonal
polynomials are badly convergent~\cite{CZ,JBJB} requesting high-precision
representations~\cite{JBJB} which are difficult to handle in practical
applications.

Another method of solution
 consists in a forming also Mellin moments in
the variable $z_-$,
 $\YINT dz_- z_-^k G^i_n(z_-) = G^i_{n k}$. Since the kernels
${\Gamma}_n^{i,j}(z_-,z'_-)$ and
$\Delta{\Gamma}_n^{i,j}(z_-,z'_-)$  obey the representation
\begin{eqnarray}
 {\Gamma}_n^{i,j}(z_-,z'_-)= \int_0^1 dw_2
 w_2^{n-h_j}
\hat K^{i,j}(z_--w_2z'_-,w_2).
\end{eqnarray}
the evolution equations may be written as (see also \cite{BGR})
\begin{eqnarray}
\label{gnk}
\mu^2 \frac{d}{d \mu^2}G_{n k}^i=
 \frac{\alpha_s(\mu^2)}{2 \pi}\sum_{j,l}
 {\Gamma}_{n, k l}^{i,j}G_{n l}^j~.
\end{eqnarray}
with the transformed kernels given by
\begin{eqnarray}
 {\Gamma}_{n, k l}^{i,j}= \frac{1}{2}\int
Dw~w_2^{n+l-h_j}w_1^{k-l}
\hat K^{i,j}(w_1,w_2)\left( \begin{array}{c}
 k \\ l \end{array} \right )~.
\end{eqnarray}
These evolution equations are not diagonal with respect to the
indices
$(k,l)$.
The explicit expressions for
${\Gamma}_{n, k l}^{i,j}$ and
$\Delta{\Gamma}_{n, k l}^{i,j}$ are given in ref.~[7,5b].
For fixed $n$
they form triangular matrices. The eigenvalues are
the diagonal elements $k=l$
\begin{eqnarray}
 {\Gamma}_{n, k k}^{i,j}= \frac{1}{2}\int Dw~w_2^{n+k-h_j}
\hat K^{i,j}(w_1,w_2)= \gamma^{i,j}_{n+k-h_j}~.
\end{eqnarray}
The coefficients $\gamma^{i,j}_{n+k-h_j}$
are the anomalous dimensions of the forward case with a shifted
Mellin index.
\section{Conclusions}
The evolution kernels for the twist~2 light--ray operators both
for the case of unpolarized and polarized deep inelastic non--forward
scattering were derived for the flavor non--singlet and singlet
cases. In general  the partition  functions depend on two distribution
variables. One may study as well specialized evolution equations in one
distribution variable implying external  constraints, covering the case
of evolution equations for  non--forward parton densities. In this way,
among various others, also the well--known evolution equations as the
Brodsky--Lepage or Altarelli--Parisi equations can be obtained.
The solution of the two-variable evolution equations in the non--singlet
and singlet cases can be either performed applying a single Mellin
transform and using Gegenbauer polynomials or by a two-fold Mellin
transform.

\vspace{3mm}
\noindent
{\bf Acknowledgement}~We would like to thank Paul S\"oding for
his constant support of the  project and H.-J.Kaiser, D. M\"uller,
A.~V.~Radyushkin and  I.~I.~Balitskii for discussions on the present topic.
J.B. would like to thank James F. Botts for many discussions on the
problem of high-precision orthogonal polynomial representations of
evolution kernels.


\end{document}